\documentclass[twocolumn,showpacs,superscriptaddress,amsmath,amssymb]{revtex4}
\usepackage{graphicx}
\usepackage{dcolumn}
\usepackage{bm}
\usepackage{color}

\def\bd{
\begin{document}} \def\ed{\end{document}}
\def\bmp{\begin{minipage}} \def\emp{\end{minipage}}
\def\bcc{\begin{center}} \def\ecc{\end{center}}     \def\npg{\newpage}
\def\beq{\begin{equation}} \def\eeq{\end{equation}} \def\hph{\hphantom}
\def\be{\begin{equation}} \def\ee{\end{equation}} \def\r#1{$^{[#1]}$}
\def\n{\noindent} \def\ni{\noindent} \def\pa{\parindent}
\def\hs{\hskip} \def\vs{\vskip} \def\hf{\hfill} \def\ej{\vfill\eject}
\def\cl{\centerline} \def\ob{\obeylines}  \def\ls{\leftskip}
\def\underbar#1{$\setbox0=\hbox{#1} \dp0=1.5pt \mathsurround=0pt
   \underline{\box0}$}   \def\ub{\underbar}    \def\ul{\underline}
\def\f{\left} \def\g{\right} \def\e{{\rm e}} \def\o{\over} \def\d{{\rm d}}
\def\vf{\varphi} \def\pl{\partial} \def\cov{{\rm cov}} \def\ch{{\rm ch}}
\def\la{\langle} \def\ra{\rangle} \def\EE{e$^+$e$^-$} \def\pt{p_{\rm t}}
\def\pti{p_{{\rm t},i}} \def\vti{v_{{\rm t},i}}
\def\ptj{p_{{\rm t},j}}\def\Pt{P_{\rm t}} \def\vt{v_{\rm t}}
\def\YT{Y_{{\rm T}}} \def\yT{y_{{\rm T}}}
\def\yTi{y_{{\rm T},i}}

\def\bitz{\begin{itemize}} \def\eitz{\end{itemize}}
\def\btbl{\begin{tabular}} \def\etbl{\end{tabular}}
\def\btbb{\begin{tabbing}} \def\etbb{\end{tabbing}}
\def\beqar{\begin{eqnarray}} \def\eeqar{\end{eqnarray}}
\def\\{\hfill\break} \def\dit{\item{-}} \def\i{\item}
\def\bbb{\begin{thebibliography}{9} \itemsep=-1mm}
\def\ebb{\end{thebibliography}} \def\bb{\bibitem}
\def\bpic{\begin{picture}(260,240)} \def\epic{\end{picture}}
\def\akgt{\cl{\bf ACKNOWLEDGMENTS}}
\def\fgn{\noindent{\bf\large\bf figure captions}}
\def\m1{\langle N_p\rangle} \def\u2{\langle N_{\bar p}\rangle} \def\Nap{N_{\bar
p}}
\def\lan{\langle}
\def\ran{\rangle}
\def\p{\pi}
\def\ifmath#1{\relax\ifmmode #1\else $#1$\fi}%
\def\rc{\ifmath{{\mathrm{c}}}}
\def\cut{\ifmath{{\mathrm{cut}}}}
\def\rF{\ifmath{{\mathrm{F}}}}
\def\rK{\ifmath{{\mathrm{K}}}}
\def\rp{\ifmath{{\mathrm{p}}}}
\def\rt{\ifmath{{\mathrm{t}}}}
\def\LAB{\ifmath{{\mathrm{LAB}}}}
\def\cut{\ifmath{{\mathrm{cut}}}}
\def\beq{\begin{equation}}
\def\eeq{\end{equation}}

\newcommand{\cinst}[2]{$^{\mathrm{#1}}$~#2\par}
\newcommand{\crefi}[1]{$^{\mathrm{#1}}$}
\newcommand{\crefii}[2]{$^{\mathrm{#1,#2}}$}
\newcommand{\crefiii}[3]{$^{\mathrm{#1,#2,#3}}$}
\newcommand{\HRule}{\rule{0.5\linewidth}{0.5mm}}

\bd
\title{Azimuthal distribution of mean transverse rapidity in relativistic heavy ion collisions}

\author{Peng Yang} 
\affiliation{Key Laboratory of Quark and Lepton Physics (MOE) and Institute of Particle Physics, Central China Normal University, Wuhan 430079, China}
\author{Lin Li} 
\affiliation{Key Laboratory of Quark and Lepton Physics (MOE) and Institute of Particle Physics, Central China Normal University, Wuhan 430079, China}
\author{Yuanfang Wu} 
\affiliation{Key Laboratory of Quark and Lepton Physics (MOE) and Institute of Particle Physics, Central China Normal University, Wuhan 430079, China}

\begin{abstract}
By the sample generated by the AMPT with string melting model, we compare the azimuthal distributions of multiplicity, total transverse rapidity, and mean transverse rapidity. It is demonstrated that the azimuthal distribution of mean transverse rapidity is a good probe of the radial kinetic expansion. The anisotropic part of the distribution characterizes the anisotropic nature of the radial expansion, and isotropic part is combinations of thermal motion and isotropic radial expansion.
\end{abstract}


\maketitle
\section{Introduction}

The study of anisotropic collective flow in relativistic heavy ion collisions has attracted increased attention. In non-central collisions, the overlap area of two incident nuclei is an almond shape in the transverse coordinate plane~\cite{lab1}. This initial geometric asymmetry causes a larger density gradient along the short axis. It finally results in the anisotropy of density distribution in momentum space. Therefore, the azimuthal distribution of final-state particles can provide valuable information about the anisotropy evolution~\cite{lab2}.

Conventionally, the azimuthal distribution of the multiplicity of final-state particles has been carefully discussed. Its Fourier expansion~\cite{lab3} is,
\begin{equation}
\label{eq1} \frac{dN}{d\phi}\propto \
1+\sum_{n=1}^{\infty}2v_{n}(N)\cos(n\phi),
\end{equation}
where \noindent $\phi$ is the azimuthal angle between the transverse momentum of the particle and the reaction plane. The Fourier coefficients are evaluated by,
\begin{equation}
\label{eq2}  v_{n}(N)=\langle\cos(n\phi)\rangle,
\end{equation}
\noindent where $\langle$\ldots$\rangle$ is an average over all particles in all events. The second harmonic coefficient $v_2(N)$ is the so-called elliptic flow parameter. It presents the anisotropy of particle density in formed system.

As we know, the initial geometric asymmetry will lead not only the anisotropy of density distribution, but also the kinetics radial expansion. In order to measure the radial kinetics expansion, the azimuthal distribution of mean transverse momentum is suggested and compared with that of total transverse momentum~\cite{lab4}. It is shown that mean transverse momentum well presents the radial kinetics expansion, in addition to the density. However, the transverse momentum of final state particles is not convenient in Lorentz transformation. The corresponding rapidity is merely a shift in Lorentz transformation. It directly related to the transverse rapidity of a final state hadron at kinetic freeze-out~\cite{lab5} by,
\begin{equation}
\label{eq3}  y_{T}=\ln(\frac{m_{T}+p_{T}}{m_{0}}),
\end{equation}
where \noindent $m_{0}$ is the particle mass in the rest frame, $p_{T}$ is transverse momentum, and $m_{T}=\sqrt{m_{0}^2+p_{T}^2}$ is the transverse mass. So we further show the azimuthal distribution of mean transverse rapidity in the paper, and compare it with those of total transverse rapidity and multiplicity.

The paper is organized as follows. Azimuthal distributions of mean transverse rapidity, total transverse rapidity, and multiplicity of final state particles are presented and compared in section II. The centrality dependence of the anisotropic part of azimuthal distribution of mean transverse rapidity is presented and studied in section III, and mass dependence of its isotropic part is presented and discussed in session IV. Finally, the summary and conclusions are drawn in section V.

\section{Azimuthal distributions}

Usually, the total transverse rapidity is the summation of all particles¡¯ transverse rapidity in an event in the $m$th azimuthal bin, i.e.,
\begin{equation}
\label{eq4}\la\YT(\phi_{m})\ra=\frac{1}{N_{\mathrm{event}}}\sum_{j=1}^{N_{\mathrm{event}}}\left(\sum_{i=1}^{N_{m}}\yTi(\phi_m)\right),
\end{equation}
where \noindent $\yTi$ is the transverse rapidity of the $i$th particle, and $N_{m}$ is the total number of particles in the $m$th bin, and the average $\la\dots\ra$ is over all events. The mean transverse rapidity is its average over total number of particles in the $m$th azimuthal bin, i.e.,
\begin{equation}
\label{eq5} \la\la\yT(\phi_m)\ra\ra=\frac{1}{N_{\mathrm{event}}}\sum_{j=1}^{N_{\mathrm{event}}}\left(\frac{1}{N_m}\sum_{i=1}^{N_{m}}\yTi(\phi_m)\right),
\end{equation}
where the average $\la\la\dots\ra\ra$ is first over all particles in the $m$th angle bin and then over all events. It measures the mean of transverse kinetic expansion. The contribution of the number of particles is reduced by the first average.

\begin{figure*}
\includegraphics[width=2.2in]{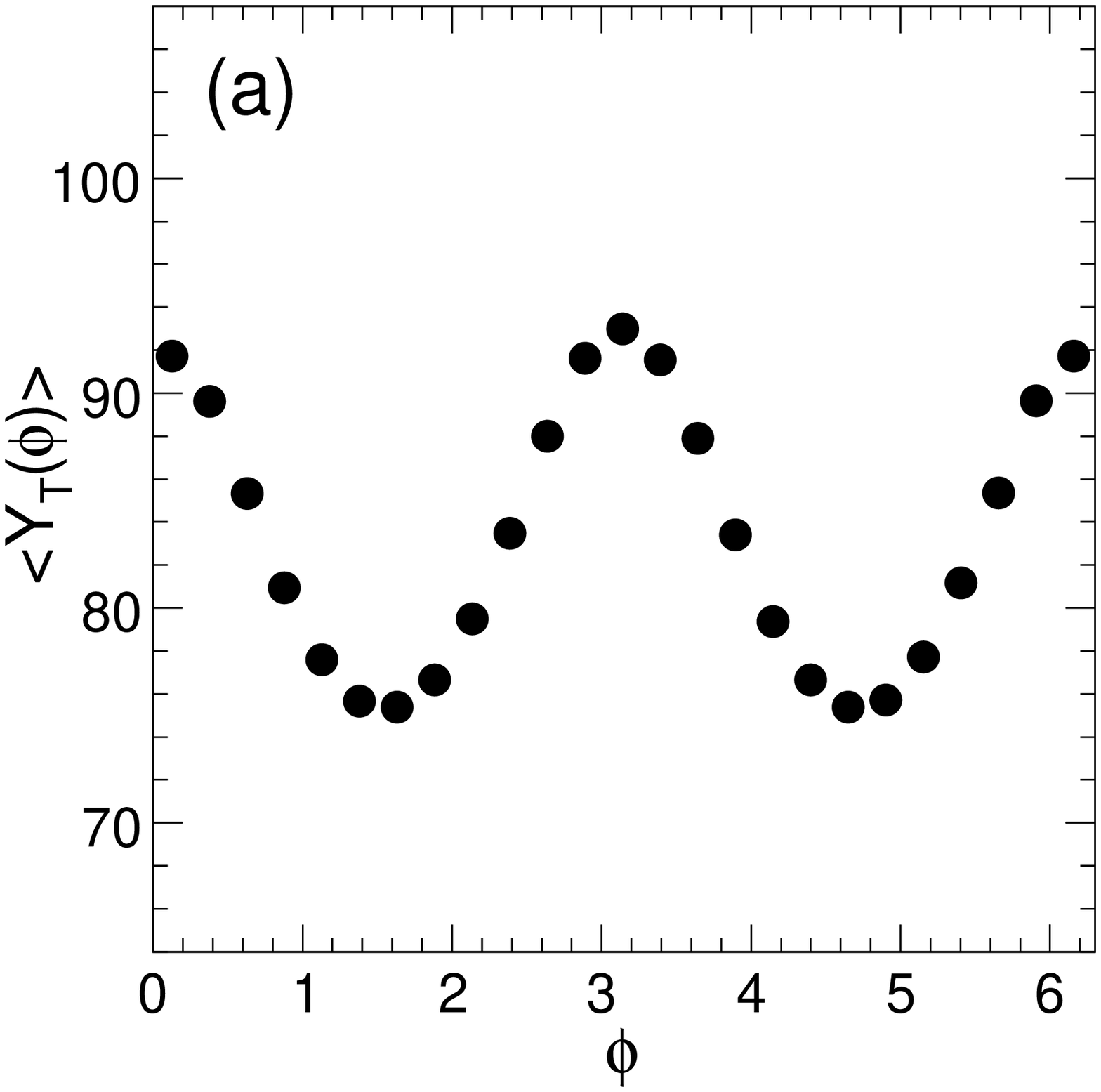}
\includegraphics[width=2.2in]{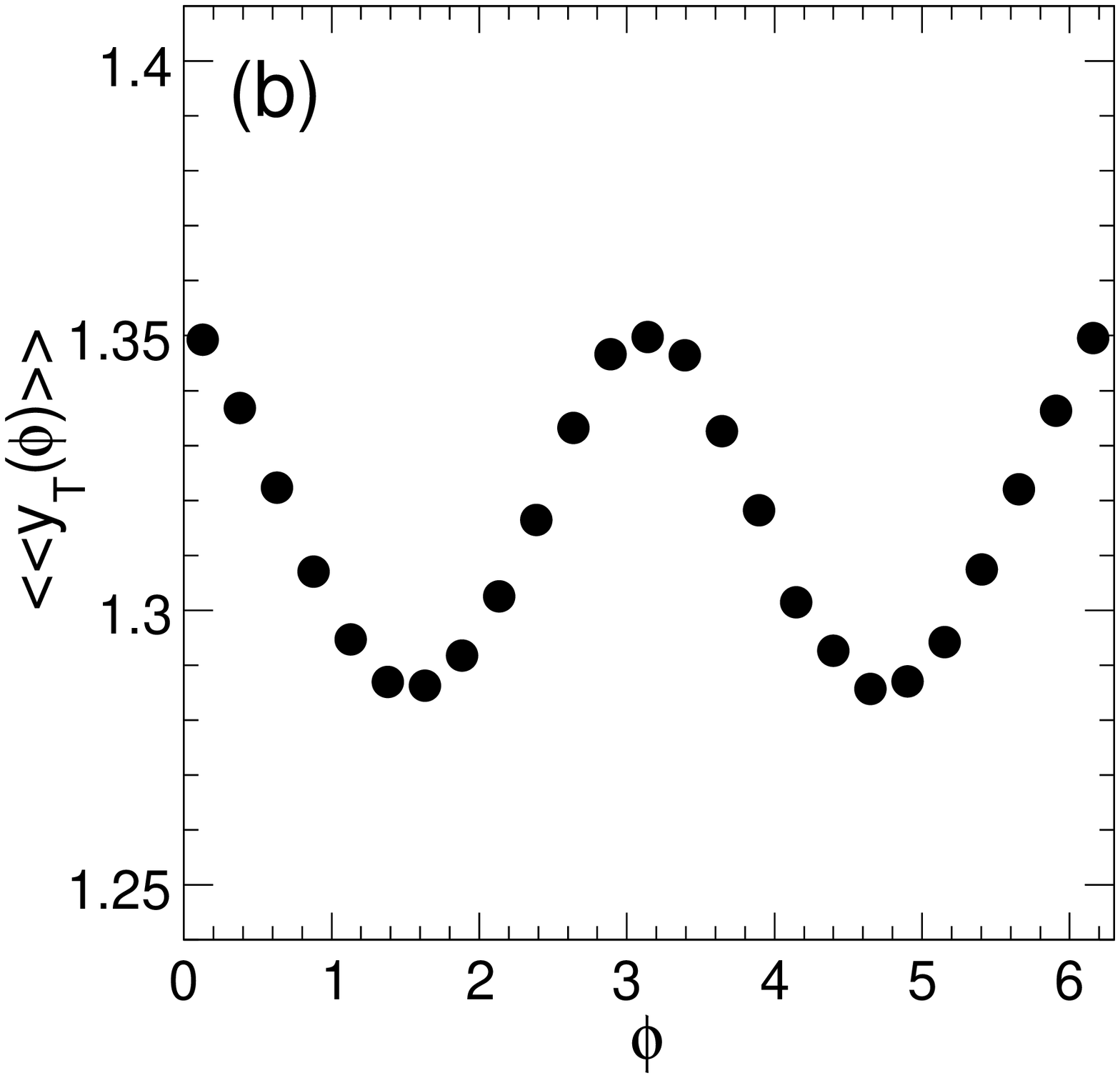}
\includegraphics[width=2.2in]{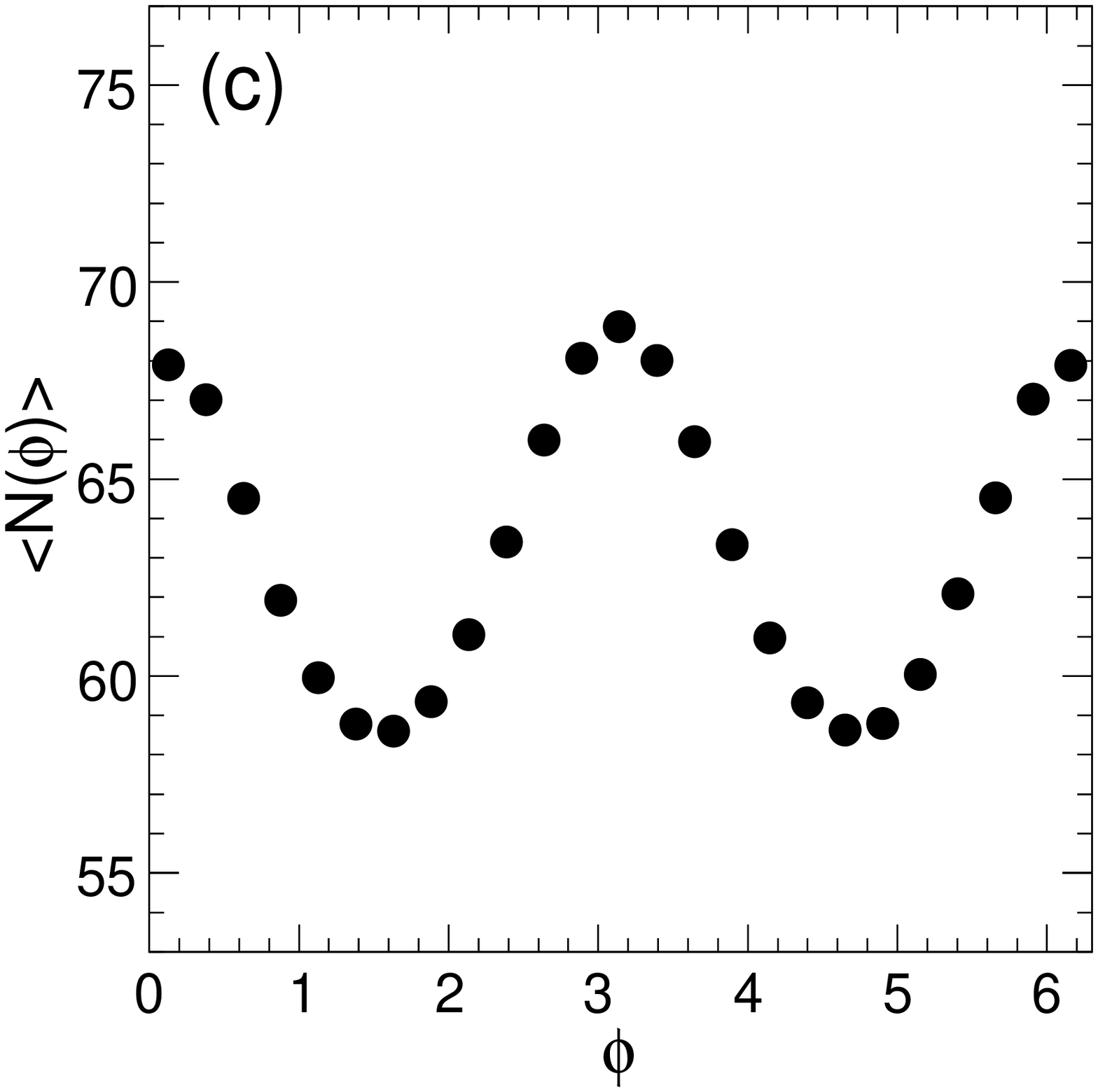}
\caption{\label{Fig. 1} The azimuthal distributions of (a) transverse rapidity, (b) mean transverse rapidity, and (c) multiplicity of minimum bias sample.}
\end{figure*}

In order to see how these distributions and what their characters are, we use the sample of Au + Au collisions at 200 GeV generated by the AMPT with string melting model~\cite{lab6,lab7} as an example to demonstrate. Where a partonic phase is implemented in the model and the elliptic flow data from RHIC are well reproduced by the model~\cite{lab8}. Here we generate about 0.2 millions minimum bias events. Azimuthal distributions of total transverse rapidity, mean transverse rapidity, and multiplicity in the full rapidity range $y\in[-5,5]$ are presented in Fig.~1(a), (b) and (c), respectively.

Although their amplitudes are quite different, they are same periodic function. Their maximums and minimums appear in the in-plane and out-plan directions, respectively. It shows that the radial expansion has the same anisotropy as that of multiplicity distribution. They can be expanded by Fourier series as well,
\begin{equation}\label{v2-YT}
 \frac{d\la\YT(\varphi)\ra}{d\varphi}=v_{0}(\la\YT\ra)[1+\sum_{n=1}^{\infty}2v_{n}(\la\YT\ra)\cos(n\varphi)],
\end{equation}
\noindent and
\begin{equation}\label{v2-yT}
 \frac{d\la\la\yT(\varphi)\ra\ra}{d\varphi}=v_{0}(\la\la\yT\ra\ra)[1+\sum_{n=1}^{\infty}2v_{n}(\la\la\yT\ra\ra)\cos(n\varphi)].
\end{equation}
Where $v_{n}(\la\YT\ra)$ and $v_{n}(\la\la\yT\ra\ra)$ are their Fourier coefficients. From the Fig.~1, we can see that the main contribution comes from the second term, $n=2$, i.e.,
\begin{equation}\label{v2-YT-fit}
\frac{d\la\YT(\varphi)\ra}{d\varphi}\approx v_{0}(\la\YT\ra)[1+2v_{2}(\la\YT\ra)\cos(2\varphi)],
\end{equation}
\noindent and
\begin{equation}\label{v2-yT-fit} \frac{d\la\la\yT(\varphi)\ra\ra}{d\varphi}\approx v_{0}(\la\la\yT\ra\ra)[1+2v_{2}(\la\la\yT\ra\ra)\cos(2\varphi)].
\end{equation}
Where the constant terms are isotropic part, and the second terms are anisotropic part. In the following two sections, we will show their physical features respectively.

\section{The features of anisotropic part of the distributions}

Since the value of anisotropic part directly relates to the centrality of the collisions, we present the centrality dependence of three kinds of anisotropic coefficients, $v_{2}(\la\YT\ra)$ , $v_{2}(N)$ , and $v_{2}(\la\la\yT\ra\ra)$ by black solid cycles, blue solid triangles and red solid stars in Fig.~2, respectively.

\begin{figure}
\includegraphics[width=3.4in]{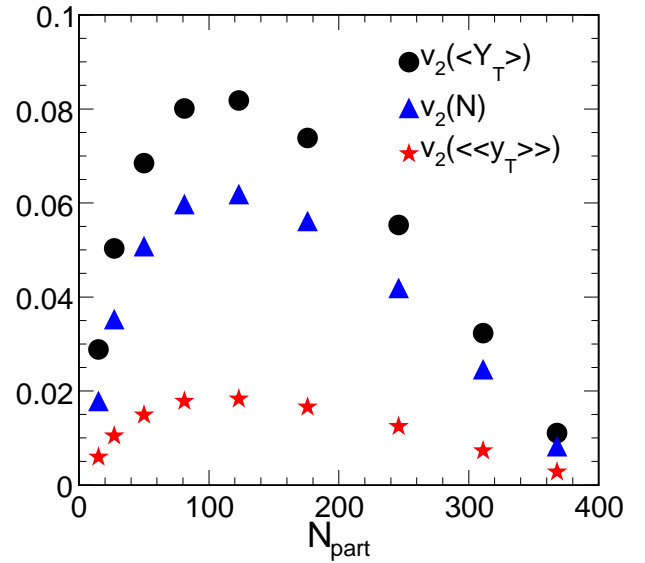}
\caption{\label{Fig. 2} The centrality dependence of anisotropic coefficients of azimuthal distributions of total transverse rapidity (solid black cycles), multiplicity (solid blue triangles) and mean transverse rapidity (solid red stars).}
\end{figure}

The figure shows that they have the similar centrality dependences. At each of centrality, the $v_{2}(\la\la\yT\ra\ra)$ is the smallest, the $v_{2}(N)$ is in the middle, and $v_{2}(\la\YT\ra)$ is the largest. This is understandable. From the definitions of Eq.~(4) and (5), it is known that the total transverse rapidity counts the information from both multiplicity and mean transverse rapidity distributions. Its anisotropy should be the largest. The mean transverse rapidity and multiplicity describe respectively the radial expansion and particle density. Their anisotropic parts are smaller than that of total transverse rapidity. The results further indicate that the anisotropy of radial kinetic expansion is smaller than that of the multiplicity distribution.

\begin{figure}[h]
\includegraphics[width=3.4in]{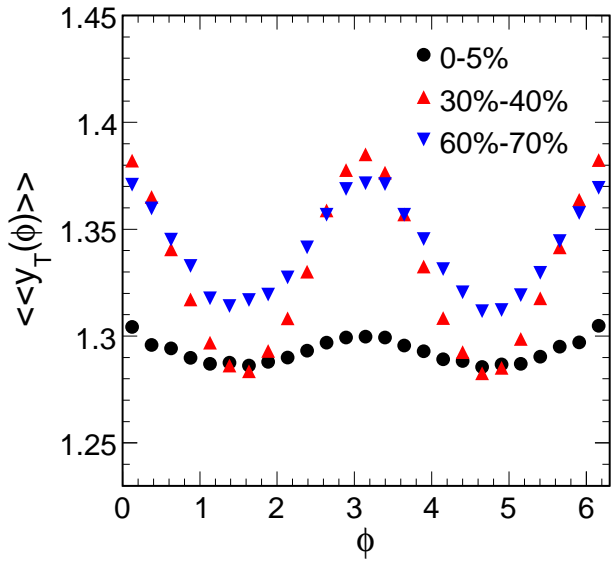}
\caption{\label{Fig. 3} The azimuthal distributions of mean transverse rapidity at three different centralities $0-5\%$ (black points), $30\%-40\%$ (red triangles), and $60\%-70\%$ (blue down triangles).}
\end{figure}

It also shows that the anisotropy of mean transverse rapidity, the red solid stars, are most close to the zero in central collisions. This means that in central collisions, the radial expansion is almost isotropic. In order to see this in more detail, the azimuthal distributions of mean transverse rapidity at three typical centralities, $0-5\%$, $30\%-40\%$, and $60\%-70\%$, are presented by black solid cycles, red solid triangles, and blue solid down-triangles in Fig.~3, respectively. Indeed, black solid cycles for central collisions are nearly azimuthal angle independent, but the red solid triangles, and the blue solid down-triangles for non-central collisions are azimuthal dependent. The large anisotropy appears in the mid-central collisions, and the small anisotropy is in the peripheral collisions. This is consistent with the expectations that anisotropic radial flow appears in non-central collisions, and is the largest in mid-central collisions~\cite{lab9}. So the anisotropic part of azimuthal distribution of mean transverse rapidity well presents the anisotropy of radial kinetic expansion.

\section{The features of isotropic part of the distribution}

The isotropic part of the azimuthal distribution of mean transverse rapidity, Eq.~(9), is a combination of the radial rapidity and thermal motion rapidity. The thermal motion is mainly determined by the temperature and particle mass. For a system at fixed temperature, the lighter particles have larger thermal velocity.

\begin{figure}[h]
\includegraphics[width=1.6in]{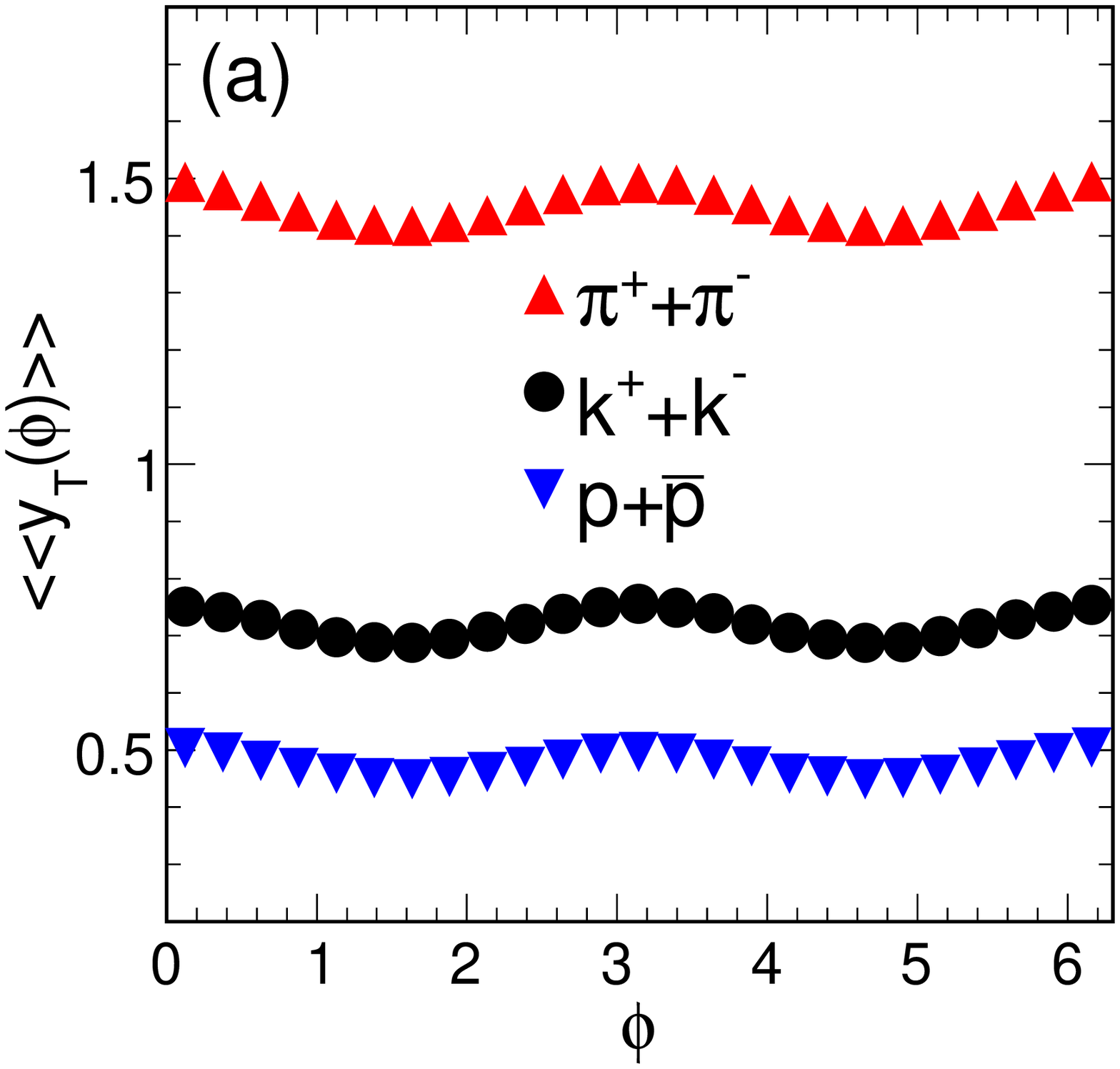}
\includegraphics[width=1.6in]{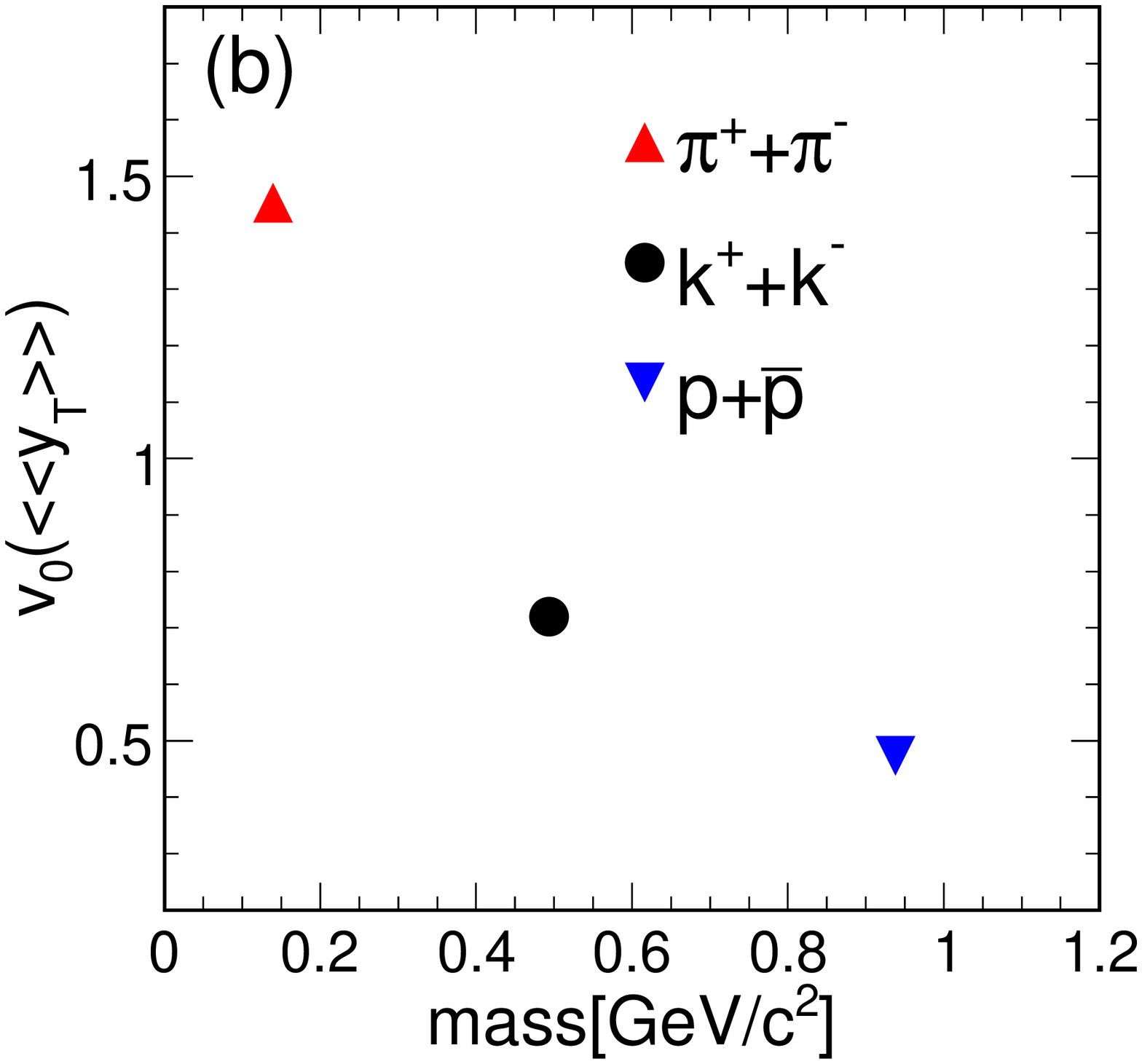}
\caption{\label{Fig. 4} (a) The azimuthal distributions of mean transverse rapidity for three different mass particles. (b) The mass dependence of the isotropic mean rapidity.}
\end{figure}

To see particle mass dependence of isotropic part, the azimuthal distributions of mean transverse rapidity for three different particle species, ¦Ð, k, and p are presented in Fig.~4(a), and the mass dependence of the isotropic parameter, $v_{0}(\la\la\yT\ra\ra)$ , is presented in Fig.~4(b). They show that the lightest particle (pion) has the largest isotropy rapidity. While the heaviest particle (proton) has the smallest isotropic rapidity, and intermediate mass particle (kaon) have the rapidity between them. These indicate their isotropic rapidities are ordered by their masses as expected random thermal motion. So the isotropic part of the azimuthal distribution of mean transverse rapidity is the combination of thermal motion and isotropic radial expansion.

\section{Summary and conclusions}

Using the sample generated by the AMPT with string melting model, we study the azimuthal distribution of mean transverse rapidity of final state particles, and compare it with the azimuthal distributions of total transverse rapidity and multiplicity. It shows that the suggested distribution mainly contains two parts: isotropic, and anisotropic mean transverse rapidity. The anisotropic part is smaller than those of total transverse rapidity and multiplicity. Its centrality dependence is as expected anisotropic radial flow. The isotropic part of the distribution is ordered by mass as expected thermal motion. It is a combination of thermal motion and isotropic radial expansion.

Therefore, the suggested distribution provides a model independent way to extract anisotropic radial rapidity. It is helpful for hydrodynamic calculations~\cite{lab10}, and a model independent determination of shear viscosity in relativistic heavy ion collisions~\cite{lab11}.

\ed